\documentclass{llncs}
\usepackage{url,graphicx,listings}
\title{Capturing Hiproofs in HOL Light\thanks{The final publication is available at \texttt{http://link.springer.com}}}
\author{Steven Obua \and Mark Adams \and David Aspinall}
\institute{School of Informatics, University of Edinburgh}
\begin{document}
\maketitle
\begin{abstract}
Hierarchical proof trees (hiproofs for short) add structure to
ordinary proof trees, by allowing portions of trees to be
hierarchically nested.  The additional structure can be
used to abstract away from details, or to label
particular portions to explain their purpose.  

In this paper we present two complementary methods for capturing
hiproofs in HOL Light, along with a tool to produce
web-based visualisations.  The first method uses \emph{tactic
  recording}, by modifying tactics to record their arguments and
construct a hierarchical tree; this allows a tactic proof script to be
modified.  The second method uses \emph{proof recording}, which
extends the HOL Light kernel to record hierachical proof trees
alongside theorems.  This method is less invasive, but requires
care to manage the size of the recorded objects.   We have
implemented both methods, resulting in two systems: \emph{Tactician}
and \emph{HipCam}.
\end{abstract}

\section{Overview}

Proofs constructed by an interactive theorem proving system can be
extremely complex.  This complexity is reflected in the size of input
proof scripts needed for large verifications, which may amount to
hundreds of thousands of lines of proof script.  It is also reflected
in attempts to demonstrate the overall result of a proof development as a
proof tree: real proof trees rapidly become large and unwieldy and
the debate over their utility continues.
%% However, they may be the best hope for exploiting the result
%% of large proofs as scripts are arbitrarily complex to understand.

A common way of managing complexity is by introducing hierarchy.  This
can be done in the input proof script language: an example is the Isar
proof language~\cite{wenzel_isar_1999}.  Isar uses block structure to
induce a hierarchy; new blocks are introduced for proof constructs
like induction and case distinction.

Hierarchy can also be used to tame large proof trees, which is our
focus in this paper.  We employ a notion of hierarchical proof known
as \emph{hiproofs}~\cite{denney_hiproofs:_2006,aspinall_tactics_2010}.
The hope is that by providing mechanisms to add hierarchy to proofs as
they are constructed, we may build proof trees that can be more easily
managed and exploited in useful ways.  With good interfaces, users may
be able to navigate proof trees comfortably, to zoom in on some detail
about how a proof proceeded, or to gain an oversight without having
expert knowledge of the source language.  Automatic tools may be
provided that take advantage of hierarchy and labelling, for example,
allowing operations for querying and transforming proofs (such as the
prototype query language in~\cite{DBLP:conf/lpar/AspinallDL12}) or
providing inputs for machine learning to investigate patterns in proof.

Block structured proof scripts and hierarchical proofs fit well
together; the latter can provide a semantics for the
former~\cite{whiteside_towards_2011}.  Here we chose to start work 
from HOL Light~\cite{_hol_????}, which does not have a hierarchical input language.
Proofs are constructed by composing tactics in the
meta-language OCaml.
So we need other ways of introducing hierarchy.
This is possible by several means: 
%% FURTHER WORK (TODO, explain it):
by transforming a previously produced proof tree, 
%% WE DO THESE THINGS:
by modifying standard tactics to produce nested labelled proofs, 
or by introducing dedicated user-level tactics.  We use
the tactic based mechanisms here.

\paragraph{Outline.} 
We will first give a quick introduction to hiproofs.  We then describe
two methods for obtaining hierarchical proofs in HOL Light.  Both work
by instrumenting the HOL Light theorem prover, but they work on
different levels of atomicity. The \emph{Tactician} tool works at the
layer of \emph{tactics} by modifying them so that proof information is
recorded in a goal tree. The \emph{HipCam} tool works at the layer of
\emph{inference rules} and modifies the HOL Light kernel so that
hiproofs are recorded in the theorem data structure, thereby extending
the proof recording approach described in~\cite{obua_importing_2006}
to also record hierarchy.  The two approaches have complementary
advantages and disadvantages; further discussion follows as they are
introduced and in the concluding section.

%The hiproofs for these examples can be inspected at \url{http://?/hiproofs}.

\section{Hierarchical Proofs}
\label{sec:hiproofs}
\begin{figure}
\begin{center}
\includegraphics[width=340px]{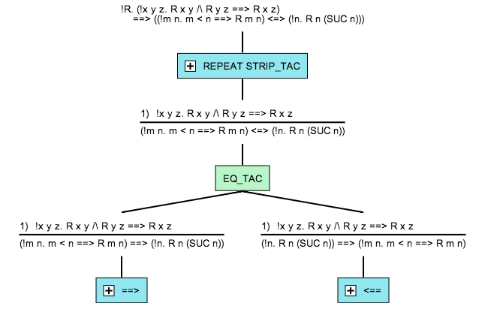}
\end{center}
\caption{Hierarchical proof of TRANSITIVE\_STEPWISE\_LT\_EQ}
\label{fig:transitive1:tactician}
\end{figure}

\begin{figure}
\begin{center}
\includegraphics[width=285px]{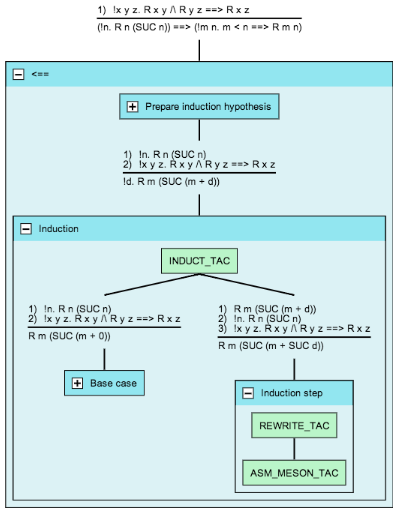}
\end{center}
\caption{Expanded box in hiproof of TRANSITIVE\_STEPWISE\_LT\_EQ}
\label{fig:transitive2:tactician}
\end{figure}

As an introductory example, Figure~\ref{fig:transitive:tactic} shows the proof of the HOL Light theorem\\
\verb+TRANSITIVE_STEPWISE_LT_EQ+.  
HOL Light is written in OCaml and therefore we use a prettified OCaml notation in this paper. Figure~\ref{fig:transitive1:tactician} shows the hierarchical proof generated from this proof by Tactician (a similar visualisation can be generated via HipCam), and Figure~\ref{fig:transitive2:tactician} shows the expanded version of the \verb+<==+ box. All boxes have been introduced automatically during the generation of the hiproof, with the exception of the box labelled ``Prepare induction hypothesis'', which has been introduced by an explicit labelling command. 

\begin{figure}
\begin{center}
\begin{verbatim}
let TRANSITIVE_STEPWISE_LT_EQ = prove
 (`!R. (!x y z. R x y /\ R y z ==> R x z)
 ==> ((!m n. m < n ==> R m n) <=> (!n. R n (SUC n)))`,
     REPEAT STRIP_TAC THEN EQ_TAC THEN ASM_SIMP_TAC[LT] THEN
     DISCH_TAC THEN SIMP_TAC[LT_EXISTS; LEFT_IMP_EXISTS_THM] THEN
     GEN_TAC THEN ONCE_REWRITE_TAC[SWAP_FORALL_THM] THEN
     REWRITE_TAC[LEFT_FORALL_IMP_THM; EXISTS_REFL; ADD_CLAUSES] THEN
     INDUCT_TAC THEN REWRITE_TAC[ADD_CLAUSES] THEN ASM_MESON_TAC[]);;
\end{verbatim}
\end{center}
\caption{Tactic-style proof of TRANSITIVE\_STEPWISE\_LT\_EQ}
\label{fig:transitive:tactic}
\end{figure}

\begin{figure}
\begin{center}
\includegraphics{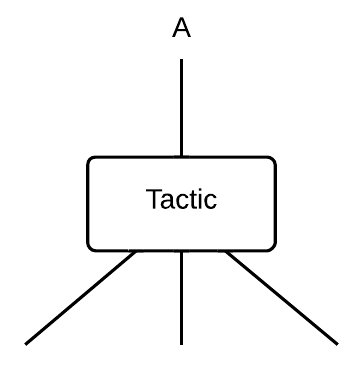}
\end{center}
\caption{$\texttt{Atomic}\,(\textrm{Tactic}, A, 3)$ }
\label{fig:basichiproof}
\end{figure}

\begin{figure}
\begin{center}
\includegraphics[scale=0.8]{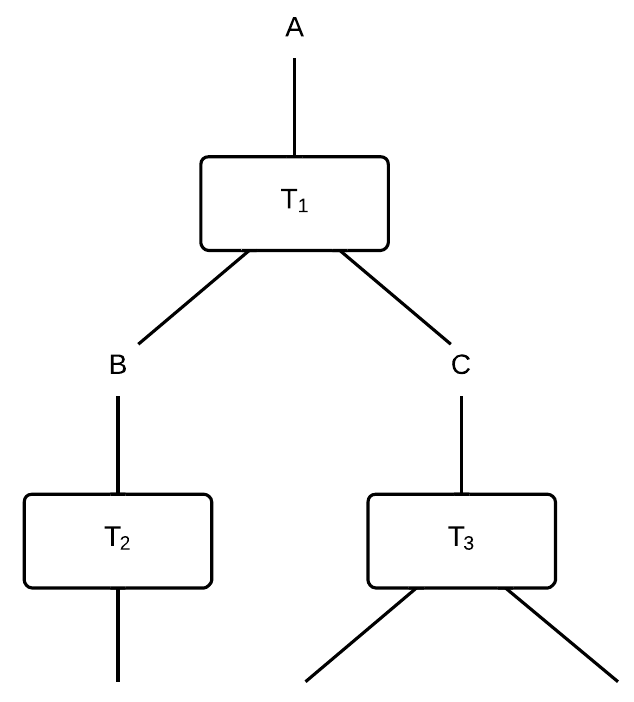}
\hspace{0.5cm}
\includegraphics[scale=0.8]{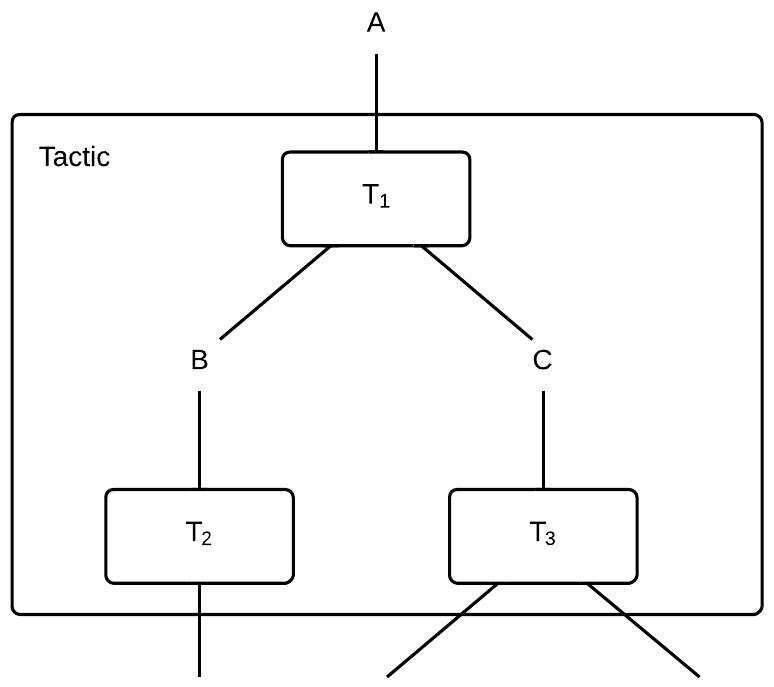}
\end{center}
\caption{Composite hiproof (left) and its boxed up version (right)}
%\caption{$\texttt{Sequence}\,[\texttt{Atomic}\,(T_1, A, 2),\,\texttt{Tensor}\,[\texttt{Atomic}\,(T_2, B, 1),\,\texttt{Atomic}\,(T_3, C, 2)]]$ }
\label{fig:complexhiproof}
\end{figure}

Hiproofs were introduced by Denney et al~\cite{denney_hiproofs:_2006},
as a uniform formalisation of ideas that had been experimented with in
several proof development systems.  Denotationally,
hiproofs are described as a forest of trees with a nesting relation.
A syntactic formulation was added later~\cite{aspinall_tactics_2010};
adapted to the purposes of this paper, this syntax can be represented
as a datatype as follows:
\begin{lstlisting}
type hiproof = 
        Atomic of label $\times$ goal $\times$ int
      | Sequence of hiproof list
      | Tensor of hiproof list
      | Box of label $\times$ hiproof
\end{lstlisting}
Here $\texttt{Atomic}\,(l, g, n)$ represents the application of an atomic tactic labelled $l$ to a goal $g$ yielding $n$ subgoals (Fig.~\ref{fig:basichiproof}) whereas
\texttt{Sequence} and \texttt{Tensor} are used to build more complex proofs. The left hiproof in Figure~\ref{fig:complexhiproof} illustrates this, the picture shown corresponds to the hiproof expression $g$ defined by \[g = \texttt{Sequence}\,[\texttt{Atomic}\,(T_1, A, 2),\,\texttt{Tensor}\,[\texttt{Atomic}\,(T_2, B, 1),\,\texttt{Atomic}\,(T_3, C, 2)]].\] The basic idea of hierarchical proofs is now that tactics are not necessarily atomic but that it is possible to ``look inside'' a tactic by representing its inside as a hiproof, too. The expression $\texttt{Box}\,(l, h)$ allows this and denotes a tactic labelled $l$ with an inner hiproof. The right hiproof in Figure~\ref{fig:complexhiproof} boxes up the hiproof $g$ to its left and is written in our notation as  $\texttt{Box}\,(\textrm{``Tactic''},\, g)$. 

Labels are arbitrary and can be used for different purposes; they can contain simple names for tactics or proof methods as we show in examples, or could, for example, contain references into the source code of the proof.  

% Our implementations are generic so that users can generate new kinds of labels that suit their particular needs. 

We are only interested in \emph{well-formed} hiproofs. In well-formed hiproofs sequences and tensors are at least two elements long. 
To check further requirements, let the function \texttt{IN : hiproof $\rightarrow$ int} be defined via
\begin{lstlisting}
  IN(Atomic($l$,$g$,$n$)) = 1
  IN(Sequence[$e$,$\ldots$]) = IN($e$)
  IN(Tensor[$e_1$,$\ldots$,$e_n$]) = IN($e_1$) + $\ldots$ + IN($e_n$)
  IN(Box($l$,$h$)) = IN($h$)
\end{lstlisting}
and let the function  \texttt{OUT : hiproof $\rightarrow$ int} be defined via
\begin{lstlisting}
  OUT(Atomic($l$,$g$,$n$)) = $n$
  OUT(Sequence[$\ldots$,$e$]) = OUT($e$)
  OUT(Tensor[$e_1$,$\ldots$,$e_n$]) = OUT($e_1$) + $\ldots$ + OUT($e_n$)
  OUT(Box($l$,$h$)) = OUT($h$)
\end{lstlisting}
Now \texttt{IN($h$)} denotes the number of subgoals that the proof $h$ proves, and \texttt{OUT($h$)} is the number of subgoals that still need to be proved after $h$ has been considered. Then a well-formed hiproof $H$ is subject to the following additional constraints:
\begin{itemize}
\item Every hiproof contained in $H$ is well-formed, too.
\item $H$ = \texttt{Box}($l$, $h$) implies \texttt{IN}($h$) = 1.
\item $H$ = \texttt{Sequence}[$e_1$, $\ldots$, $e_n$] implies \texttt{OUT}($e_i$) = \texttt{IN}($e_{i+1}$) for $1 \leq i < n$.
\end{itemize}
The well-formedness constraints ensure that hiproofs are ``plugged
together'' correctly, and serve as (informal) invariants maintained in
our software.

Our implementations include a module (based on Javascript and HTML5 Canvas) that
displays well-formed hiproofs in a web browser as
shown in Figure~\ref{fig:transitive1:tactician} and
Figure~\ref{fig:transitive2:tactician}.  Boxes can be collapsed so that
they display only their label and not their inner hiproof. The display
of intermediate goals can be toggled individually.  

In the next two sections we will present two methods for capturing
hiproofs of HOL Light theorems.

\section{Tactician}

%% DA: we need to change this because it doesn't flow with the main story
%% I've added next paragraph
Tactician is a productivity tool for refactoring individual HOL Light
tactic proof scripts.  It supports two main refactoring operations:
packaging up a series of tactic steps into a single compound tactic
joined by THEN and THENL tacticals, and the reverse operation, for
flattening out a packaged-up tactic into a series of tactic steps.  It
is aimed at helping experts maintain their proof scripts, and helping
beginners learn from existing proof scripts.  It can be obtained
from~\cite{adams_tactician_2012}.

Behind the scenes, Tactician uses a representation of the recorded
tactic proof tree which is close to a hiproof; recording hierachical
proofs was one of its original design goals.
%% HILABEL mention here or below "various wrapper functions"

\subsection{Example}
\begin{figure}[t]
\begin{tabular}{ll}
\textbf{label} &\ \textbf{flattened proof}\\\hline
\begin{minipage}[t]{1cm}
\begin{verbatim}


A
B

C

D
E
F
G
H
I
J

K
L

M
N
\end{verbatim}
\end{minipage}
&\ \begin{minipage}[t]{10.8cm}
\begin{verbatim}
g `!R. (!x y z. R x y /\ R y z ==> R x z)
         ==> ((!m n. m < n ==> R m n) <=> (!n. R n (SUC n)))`;;
e (REPEAT STRIP_TAC);;
e (EQ_TAC);;
(* *** Subgoal 1 *** *)
e (ASM_SIMP_TAC [LT]);;
(* *** Subgoal 2 *** *)
e (ASM_SIMP_TAC [LT]);;
e (DISCH_TAC);;
e (SIMP_TAC [LT_EXISTS; LEFT_IMP_EXISTS_THM]);;
e (GEN_TAC);;
e (ONCE_REWRITE_TAC [SWAP_FORALL_THM]);;
e (REWRITE_TAC [LEFT_FORALL_IMP_THM; EXISTS_REFL; ADD_CLAUSES]);;
e (INDUCT_TAC);;
(* *** Subgoal 2.1 *** *)
e (REWRITE_TAC [ADD_CLAUSES]);;
e (ASM_MESON_TAC []);;
(* *** Subgoal 2.2 *** *)
e (REWRITE_TAC [ADD_CLAUSES]);;
e (ASM_MESON_TAC []);;
\end{verbatim}
\end{minipage}
\end{tabular}
\caption{Flattened proof of TRANSITIVE\_STEPWISE\_LT\_EQ}
\label{fig:transitive:flattened}
\end{figure}
A typical packaged up proof has already been presented in Fig.~\ref{fig:transitive:tactic}. The result of flattening out this proof is shown in Fig.~\ref{fig:transitive:flattened}. The following hiproof can be directly read off from the flattened proof (we omit the goals in \texttt{Atomic}):
\begin{verbatim}
Sequence[
    Atomic(A,1),Atomic(B,2), 
    Tensor[
        Atomic(C,0),
        Sequence[
            Atomic(D,1),Atomic(E,1),Atomic(F,1),
            Atomic(G,1),Atomic(H,1),Atomic(J,2),        
            Tensor[
                Sequence[Atomic(K,1),Atomic(L,0)],
                Sequence[Atomic(M,1),Atomic(N,0)]]]]]           
\end{verbatim}
This hiproof corresponds (after introduction of several \texttt{Box}es) to the
visualisation shown in Fig.~\ref{fig:transitive1:tactician} and Fig.~\ref{fig:transitive2:tactician}. 

\subsection{Tactic Recording}

% change this sentence because one ref asked for recall of classic HOL,
% but it is already here
% Typically the user interacts with the HOL Light system as follows. 
It helps to recall how a tactic proof is constructed in HOL Light.
The user starts with a single main goal, which gets broken down over a
series of tactic steps into hopefully simpler-to-prove subgoals.  The
user works on each subgoal in turn.
% , moving onto the next when the
% current subgoal has been proved.  
The proof is complete when the last
subgoal has been proved.  Behind the scenes, the standard subgoal package
maintains a proof state that consists of a list of current proof goals
and a justification function for constructing the formal proof of a
goal from the formal proofs of its subgoals.  Tactics are functions
that take a goal and return a subgoal list plus a justification
function.  The subgoal package state is updated every time a tactic is
applied, incorporating the tactic's resulting subgoals and
justification function.

Tactician works by recording such a tactic-style proof in a proof
tree, where each node in the tree corresponds to a goal in the proof.
When a user wants to refactor the proof, the proof tree is abstracted to
a hiproof, which then gets refactored accordingly before being emitted
as an ML tactic proof script.  We give a brief overview of the
recording mechanism here; more details are 
in~\cite{adams_recording_2012}.

The proof tree gets initialised when a tactic proof is started, and is added to as tactics are executed.  Tactics are modified so that they work on a modified, or ``promoted'', goal datatype called \texttt{xgoal} (Fig.~\ref{fig:holmods}).  Each \texttt{xgoal} carries a unique goal id, which corresponds to a node in the proof tree.
\begin{figure}[b]
\begin{lstlisting}
type goalid = int
type xgoal = goal $\times$ goalid
type xgoalstate = (term list $\times$ instantiation) $\times$ 
                  xgoal list $\times$ justification
type xtactic = xgoal $\rightarrow$ xgoalstate
\end{lstlisting}
\caption{Modifications of HOL Light's datatypes}
\label{fig:holmods}
\end{figure}
A modified tactic has type \texttt{xtactic}. It takes an \texttt{xgoal} input, strips away its id, applies the original unmodified tactic, and generates new ids for each of the resulting goals.  Information about the tactic step, including an abstraction of the text of the tactic as it would appear in the proof script, is then inserted into the proof tree at the node indicated by the input's id.  An index of ids and references to their corresponding nodes is maintained to enable nodes to be located.

Boxes around tactics can be introduced manually via a function
\begin{lstlisting}
val hilabel : label $\rightarrow$ xtactic $\rightarrow$ xtactic
\end{lstlisting}
so $\texttt{hilabel}(l,t)$ sets up a new box with the label $l$ around the tactic $t$.   The input of $t$ becomes the input of the box, and the subgoals which result from applying $t$ become the outputs.

Apart from basic tactics, there are tactic-producing functions which depend on additional arguments like terms, theorems, or other tactics. We also need to modify these more complicated tactic forms. Because each tactic form has a fixed ML type signature, a generic wrapper function can be written for performing this modification for each such form.  About 20 such wrappers need to be written to cover the commonly used tactic forms in the HOL Light base system.

\subsection{Capturing Hiproofs with Tactician}

Based on its tactic recording mechanism, Tactician can generate hiproofs from tactic style proofs by a straightforward transformation of the tactic proof tree to hiproofs. Because the proof tree corresponds naturally to the user's actual proof script, so does the hiproof. Hierarchical boxes can optionally be introduced by the various wrapper functions. This method of generating hiproofs works also for proofs that have not been completed yet, and can therefore potentially be used to visualise the current proof state during interactive proof as a hiproof. 

Tactician outputs a proof at the user level, i.e., involving the same atomic ML tactics, rules and theorems as occur in the original proof script.  Low-level information of the proof is not retained. This has the advantage that hierarchical proofs are maintained at a level meaningful to the user, and the overhead of recording is kept low.

One fundamental limitation of Tactician is that tactics that take functions as arguments cannot be ``promoted'' if the function itself does not return a promotable datatype (the only common instance of this in HOL Light is \verb+PART_MATCH+, which takes a term transformation function as an argument).  Another is that ML type annotations in the proof script need to mention promoted, rather than unpromoted, ML datatypes.

With Tactician version 2.2, proof script files involving several hundred lines of ML will typically encounter one or two occurrences of such limitations.  It is possible to get around these problems by making hand edits to the proof script, but highly-automated processing of very large bodies of proof is not currently feasible.

%% da: move back from this as I think the presentation of the two tools as 
%% complementary is more friendly
% Because of these difficulties we looked for an alternative way of capturing hiproofs. We will present such an alternative in the next section. 

\section{HipCam}

The basic idea of HipCam is to modify the HOL Light kernel, instead of
modifying the higher-level tactic-layer like Tactician does.  While
Tactician relies on \emph{tactic recording}, HipCam instead uses a
\emph{proof recording} approach closely related to that pioneered
in~\cite{obua_importing_2006} .  HipCam can be downloaded
from~\cite{obua_hipcam_2013}.

HipCam is minimally invasive; any theorem proven using the original
HOL Light kernel can be proven using the HipCam-modified kernel
and modification of proof scripts is not needed, except to add
explicit hierarchy and labelling, incrementally as desired.
However, HipCam does not allow recovering proof scripts from recorded
proofs; it is intended primarily as a tool to construct large hiproofs
for inspection, rather than replay or refactor proof scripts.

\subsection{Proof Recording}

HipCam does not alter the signature of the HOL Light kernel except
to add two functions.  To extract a hierarchical
proof from such a theorem in the modified kernel, one applies the new
kernel function
\begin{lstlisting}
val hiproof : thm $\rightarrow$ hiproof
\end{lstlisting}
to the theorem.  
This is made possible by changing the definition for the type \texttt{thm} from
\begin{lstlisting}
type thm = Sequent of term list $\times$ term
\end{lstlisting}
which stores the assumptions and conclusion of a theorem to
\begin{lstlisting}
type thm = Sequent of hiproof $\times$ term list $\times$ term
\end{lstlisting}
which in addition stores the hiproof of a theorem. This change in the implementation of \texttt{thm} is visible outside the kernel in only one way, by using native ML  equality to compare theorems.  Fortunately, after proof recording was introduced to HOL Light, native ML equality is not used to compare theorems anymore.  To test two theorems for equality, the function \texttt{equals\_thm} is called; it only compares assumptions and conclusions.

All kernel primitives which produce values of type \texttt{thm} are modified to also produce corresponding theorem-internal hiproofs.
None of those primitives introduce hierarchy, though. To produce hiproofs with an actual hierarchy we need another new kernel function, \texttt{hilabel}, which will  draw a box around an existing hierarchical proof.  What could be the signature of such a function?  Our first guess might be
\begin{lstlisting}
val hilabel$_1$ : label $\rightarrow$ hiproof $\rightarrow$ hiproof
\end{lstlisting}
which can simply be defined via
\begin{lstlisting}
let hilabel$_1$ $l$ $h$ = Box($l$,$h$)
\end{lstlisting}
The obvious problem with this is that still for no theorem $t$ will \texttt{hiproof($t$)} contain any boxes. This is simply because \texttt{hilabel$_1$} does not allow any change of the internal hiproofs of theorems. 

Our next guess might therefore be to rectify this problem as follows:
\begin{lstlisting}
val hilabel$_\texttt{thm}$ : label $\rightarrow$ thm $\rightarrow$ thm
let hilabel$_\texttt{thm}$ $l$ (Sequent($h$,$asms$,$concl$)) = 
      Sequent(Box($l$,$h$),$asms$,$concl$)
\end{lstlisting}
Unfortunately, \texttt{hilabel$_\texttt{thm}$} does not allow us to create boxes as flexibly as we want to. This is because for any theorem 
\texttt{$t$ =  Sequent($h$,$asms$,$concl$)}
the invariants \texttt{IN($h$) = $1$} and \texttt{OUT($h$) = $0$} hold.  In other words, no sub-goals can be exported from nested boxes and we could only
construct fully nested trees.  So a sub-hiproof 
$H$ 
like the ones from Fig.~\ref{fig:complexhiproof}
could not be contained in any of the hiproofs created via \texttt{hilabel$_\texttt{thm}$} because \texttt{OUT($H$) = $3$} holds. 

What we need when drawing a box onto a hiproof is the ability to specify which part of the hiproof should become part of the box, and which part should stay outside of the box. We gain this ability by drawing boxes around \emph{rules} instead of just theorems:
\begin{lstlisting}
type rule = thm list $\rightarrow$ thm
val hilabel : label $\rightarrow$ rule $\rightarrow$ rule
\end{lstlisting}
We will see in the next section how \texttt{hilabel} works and how it can be implemented.  Meanwhile, we can see that \texttt{hilabel} will satisfy all of our boxing needs. It is still trivial to box theorems:
\begin{lstlisting}
let hilabel$_\texttt{thm}$ $l$ $t$ = hilabel $l$ (fun _ $\rightarrow$ $t$) []
\end{lstlisting}
It is also straightforward how to label tactics (with a reminder of the types):
\begin{lstlisting}
type goalstate = (term list $\times$ instantiation) $\times$ 
                  goal list $\times$ justification
type tactic = goal $\rightarrow$ goalstate
val hilabel$_\texttt{tac}$ : label $\rightarrow$ tactic $\rightarrow$ tactic 
let hilabel$_\texttt{tac}$ $l$ $t$ $g$ =
  let (inst, gls, j) = $t$ $g$ in
  let k inst = hilabel $l$ (j inst)
  in (inst, gls, k) 
\end{lstlisting}
The above code reduces labelling a tactic to labelling the justification function that is obtained as the result of applying the tactic to a goal.

\subsection{Implementing \texttt{hilabel}}
\begin{figure}[t]
\hspace{0.05\textwidth}\includegraphics[width=0.4\textwidth]{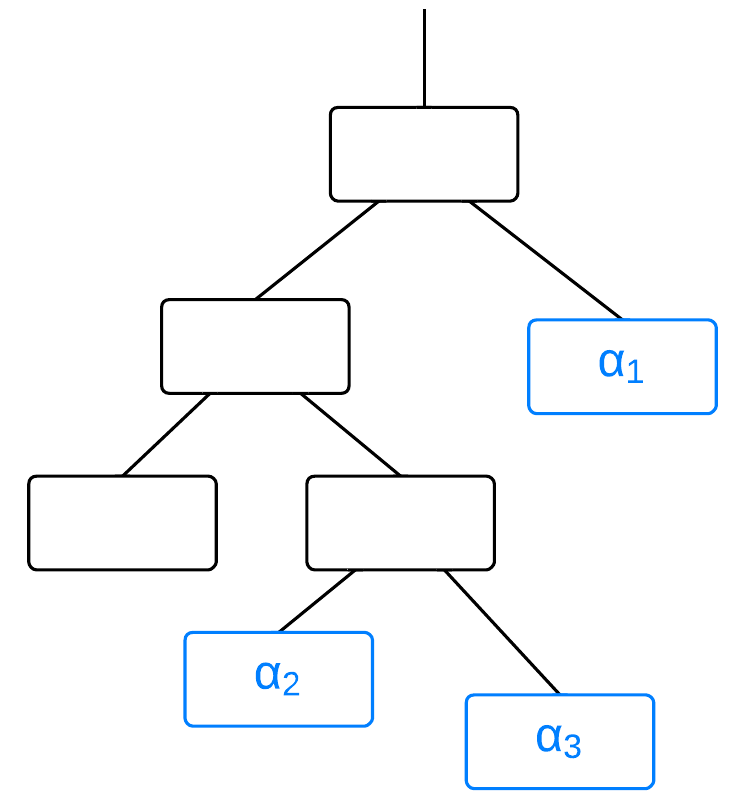}
\hspace{0.1\textwidth}\includegraphics[width=0.4\textwidth]{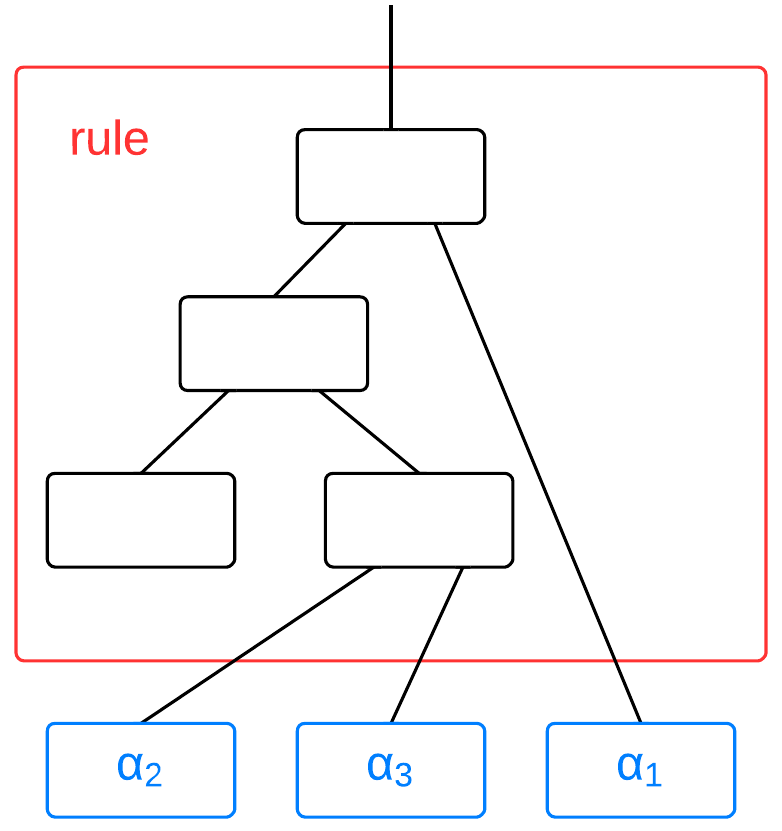}
\caption{\texttt{rule[$\alpha_1$,$\alpha_2$,$\alpha_3$]} vs. \texttt{(hilabel "rule" rule)[$\alpha_1$,$\alpha_2$,$\alpha_3$]}}
\label{fig:hilabel:alpha123}
\end{figure}

Let us examine how we want \texttt{hilabel} to behave. Assume we have a rule 
\begin{lstlisting}
val rule : thm list $\rightarrow$ thm
\end{lstlisting}
and three theorems $\alpha_1$, $\alpha_2$ and $\alpha_3$ such that 
\begin{lstlisting}
rule[$\alpha_1$,$\alpha_2$,$\alpha_3$]
\end{lstlisting}
yields a new theorem as the result of applying \texttt{rule} to these theorems. The hiproof of this new theorem will then in some way depend on the hiproofs of the 
$\alpha_i$, e.g. like depicted on the left in Fig.~\ref{fig:hilabel:alpha123}. Now, if instead of applying the original rule, we apply the labelled rule 
\begin{lstlisting}
(hilabel "rule" rule)[$\alpha_1$,$\alpha_2$,$\alpha_3$]
\end{lstlisting}
then we'd like the hiproof of the resulting theorem to look like depicted on the right in Fig.~\ref{fig:hilabel:alpha123}.
\begin{figure}[t]
\hspace{0.05\textwidth}\includegraphics[width=0.4\textwidth]{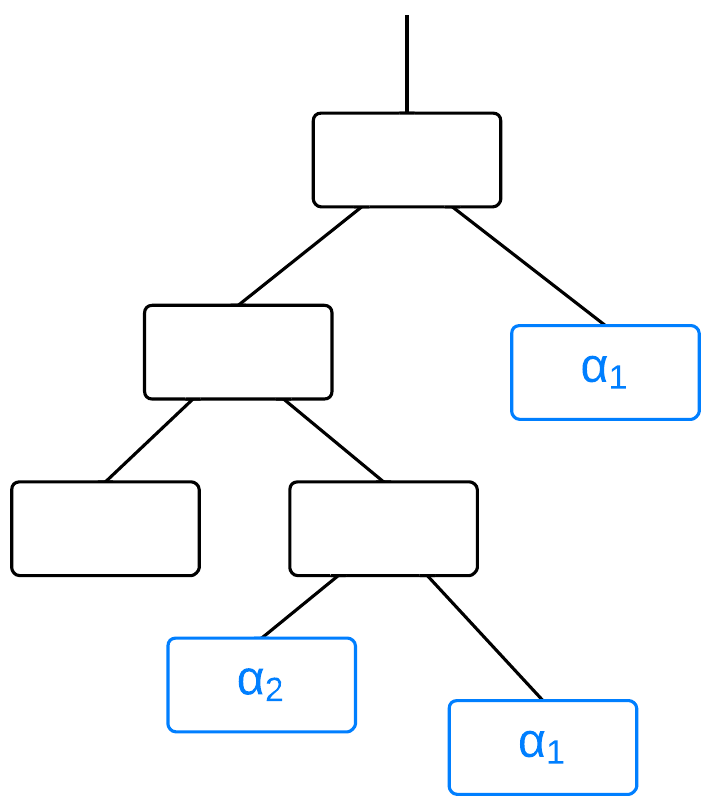}
\hspace{0.1\textwidth}\includegraphics[width=0.4\textwidth]{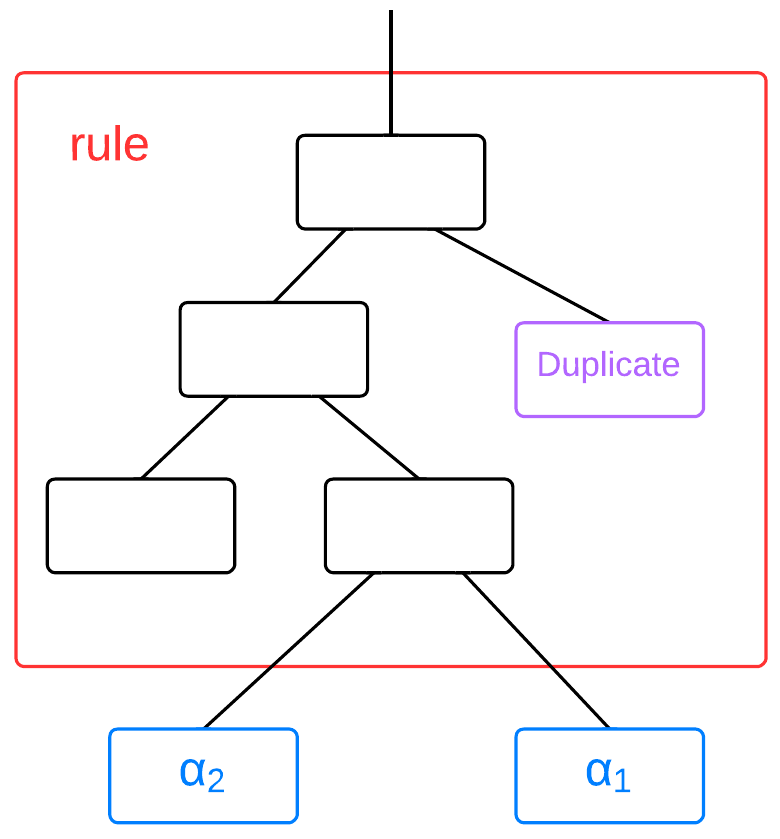}
\caption{Unused $\alpha_3$ and duplicate $\alpha_1$}
\label{fig:hilabel:alpha112}
\end{figure} 
Note that all we guarantee for the boxed hiproof is that the hiproofs of the $\alpha_i$ will appear (if at all) outside of the box. In particular, there is no predetermined order in which the hiproofs of the $\alpha_i$ will appear. It might even be the case that some of these hiproofs are not used at all, or are used more than once. This situation is shown in Fig.~\ref{fig:hilabel:alpha112}: here the proof of \texttt{rule[$\alpha_1$,$\alpha_2$,$\alpha_3$]} does not make use of 
$\alpha_3$, and uses $\alpha_1$ twice.  We detect multiple appearances of the same $\alpha_i$ and treat only the first occurrence normally. The other occurrences are marked as being duplicate instances of goals proven elsewhere. 

Our design of \texttt{hilabel} is driven by trying to make the collapsing and expanding of boxes in a visualised hiproof straightforward.  One can 
imagine other ways of dealing with reordered, duplicate, or missing dependencies. For example, we could introduce a new hiproof constructor for boxes which rewire the outputs of their inner hiproofs such that externally, the outputs of the box correspond 1-to-1 and in the right order to the arguments of the rule the box is supposed to represent (a \emph{swap} primitive can be used for this purpose;
see \cite{Whiteside:thesis}). 
% We will not explore these other possibilities further in this paper.

% Now that we have a clear idea of how we want \texttt{hilabel} to behave, let us turn towards implementing this idea. 
To implement \texttt{hilabel}, we first introduce three kinds of labels: the identity label $L_{id}$, the duplicate label $L_{dup}$ and a family of variable labels $L^{name}_{var}$ where $name$ is from some infinite set $V$ of variable names. We then define
\begin{lstlisting}
Identity($g$) $\equiv$ Atomic($L_{id}$,$g$,1)
Duplicate($g$) $\equiv$ Atomic($L_{dup}$,$g$,0)
Variable($name$,$g$) $\equiv$ Atomic($L^{name}_{var}$,$g$,0)
\end{lstlisting}
\begin{figure}[t]
\begin{center}
\includegraphics[width=0.4\textwidth]{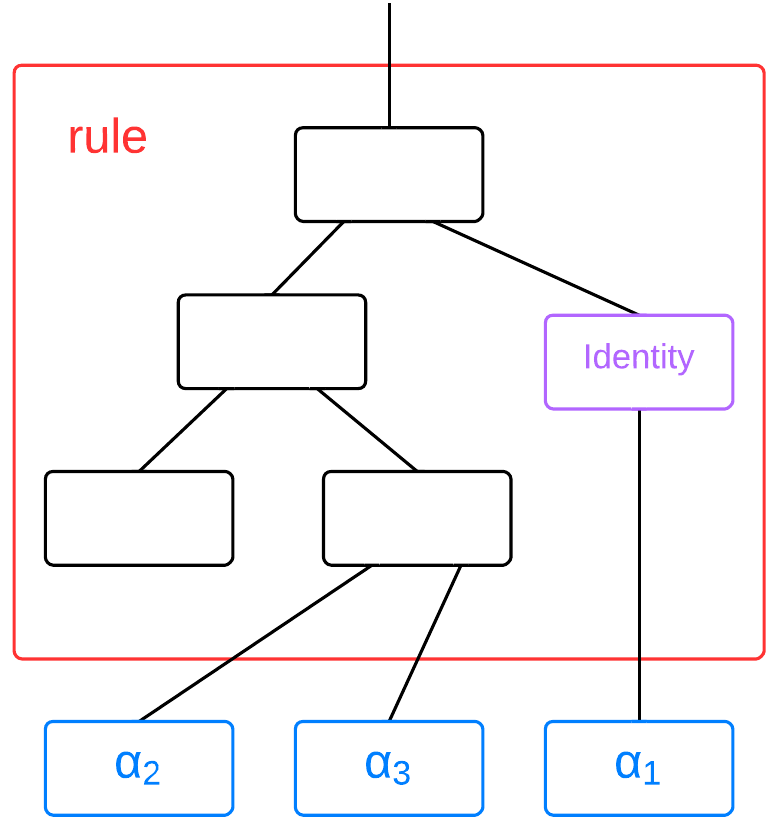}
\end{center}
\caption{Explicit display of identity tactic}
\label{fig:hilabel:identity}
\end{figure} 
to serve us as identity tactic, duplicate marker, and hiproof variable, respectively. We need the identity tactic because without it we could not represent the right hand side hiproof in Fig.~\ref{fig:hilabel:alpha123} (which is just a prettification of the hiproof shown in Fig.~\ref{fig:hilabel:identity}). We have already motivated why we need duplicate markers (Fig.~\ref{fig:hilabel:alpha112}). 
And we need variables so that we can track how the hiproofs of $\alpha_i$ are being used in constructing the hiproof for 
\texttt{rule[$\alpha_1$,$\ldots$,$\alpha_k$]}. The details of how this tracking is achieved are shown in Fig.~\ref{fig:hilabel:def}. There the notation \texttt{$\alpha$/$h$} is used to represent the theorem resulting from replacing the hiproof of the theorem $\alpha$ with the hiproof $h$. The heavy lifting in \texttt{hilabel} is done by the function $\texttt{turnvars}$.
%  which is described in Fig.~\ref{fig:turnvars}.

\begin{figure}[t]
\begin{lstlisting}
val hilabel : label $\rightarrow$ rule $\rightarrow$ thm list $\rightarrow$ thm
let hilabel $l$ $rule$ [$\alpha_1$,$\ldots$,$\alpha_k$] =
  let $N$ be a set $\{n_1,\ldots,n_k\}$ of $k$ fresh names in  
  let $\alpha'_i$ = $\alpha_i$/(Variable($n_i$, goal($\alpha_i$))), $i = 1 \ldots k$, in
  let $\beta$ = $rule$[$\alpha'_1$,$\ldots$,$\alpha'_k$] in
  let ($names$, $h$) = turnvars $N$ (hiproof($\beta$)) in
  let $H$ be a function such that $H(n_i) = \texttt{hiproof}(\alpha_i)$ in
  let $b$ = Box($l$,$h$) in
  let $h'$ = 
    match (map $H$ $names$) with
      [] $\rightarrow$ $b$
    | [$a$] $\rightarrow$ Sequence[$b$,$a$]
    | $\overline{a}$ $\rightarrow$ Sequence[$b$,Tensor($\overline{a}$)]
  in $\beta / h'$


val turnvars : $V$ set $\rightarrow$ hiproof $\rightarrow$ $V$ list $\times$ hiproof 
let turnvars $N$ $h$ =
  let $h'$ = 
      (replace all occurrences of Var($n$,$g$) in $h$ where $n \in N$ 
       either with Identity($g$) or with Duplicate($g$)
       and massage the result so that it is well-formed) 
  in let $names$ = 
      (the list of variable names which correspond 
       to the outputs of $h'$) 
  in ($names$,$h'$)
\end{lstlisting}
\caption{Description of \texttt{hilabel} and \texttt{turnvars}}
\label{fig:hilabel:def}
\end{figure}

% \begin{figure}[t]
% \begin{lstlisting}
%\end{lstlisting}
%\caption{Description of \texttt{turnvars}} 
%\label{fig:turnvars}
%\end{figure}

\newcommand{\stat}[2]{#1 / #2}

\begin{figure}[t]
\begin{center}
\begin{tabular}{l|ccc}
\textbf{Proof} & \textbf{Original} & \textbf{Max-Detail} & \textbf{High-Level}\\\hline
\texttt{\#use "hol.ml";;} & \stat{4 min 30 sec}{0.1 GB} & \stat{7 min}{2.2 GB} & \stat{6 min}{0.7 GB}\\
G\"odel 1 & \stat{10 min 30 sec}{0.1 GB} \hspace{0.1cm} & \stat{15 min}{3.1GB} & \stat{13 min}{0.9 GB}\\
$e$ is transcendental & \stat{13 min}{0.2 GB} & \stat{19 min}{4.2 GB} \hspace{0.1cm} & \stat{16 min}{1.5 GB}\\
Jordan curve theorem \hspace{0.1cm} & \stat{31 min}{0.4 GB} & out of memory & \stat{45 min}{6.5 GB}
\end{tabular}
\end{center}
\caption{HipCam statistics, using a MacBook Pro, Quad Core 2.4GHz, 16GB RAM}
\label{table:stats}
\end{figure}

A major challenge in the actual implementation of \texttt{hilabel} and
\texttt{turnvars} is that recorded proof trees quickly grow to be
enormous. Their representations in memory exploit sharing, but
repeatedly traversing such trees depth-first to compute or update them
is impractical.
%  In fact, they are so huge that even just traversing the hiproof data structure depth-first is for impractical for 
%
More sophisticated data structures %% Refs...
could help with this, but we use the simple
fix of adapting the described algorithms so that all important properties of a hiproof are computed (and then cached for shared reuse) during the construction of the hiproof, so later traversals are unnecessary. One such property of a hiproof we have introduced is its \emph{shallow size} \texttt{SS($h$)} which measures the size of a hiproof $h$ as if all boxes it contained were replaced by atomics instead:
\begin{lstlisting}
SS(Atomic($l$,$g$,$n$)) = 1
SS(Sequence[$e_1$,$\ldots$,$e_n$]) = 1 + SS($e_1$) + $\ldots$ + SS($e_n$)
SS(Tensor[$e_1$,$\ldots$,$e_n$]) = 1 + SS($e_1$) + $\ldots$ + SS($e_n$)
SS(Box($l$,$h$)) = 1
\end{lstlisting}
We can use the shallow size of a hiproof $h$ to adjust $h$ to the needs of the hiproof consumer. For visualisation, for example, we are not interested in hiproofs that have a shallow size larger than a certain threshold $\tau$, say $\tau = 1000$. Therefore we replace subexpressions of the form \texttt{Box($l$,$h$)} with an atomic whenever $\texttt{SS($h$)} > \tau$.  Note that such a replacement requires that we introduce a property for atomics which keeps track of the variables that the replaced box contained.

Another way to cut-out uninteresting detail is to look at the label
of a box.  
%
%  considerable performance gain can be achieved by performing such a replacement of \texttt{Box($l$,$h$)} not only depending on the shallow size of 
% $h$, but also depending on its label $l$. 
For example, when $l$ indicates that the box corresponds to a standard HOL Light inference rule, we could also elide the detail. Therefore HipCam has two modes: a \emph{max-detail} mode, which does not replace boxes corresponding to standard inference rules, and a  \emph{high-level} mode which does.

\subsection{Capturing Hiproofs with HipCam}

We have applied HipCam to several formalisations that ship with HOL Light. Fig.~\ref{table:stats} displays how much time and space were needed for each formalisation, without HipCam, with HipCam in max-detail mode, and with HipCam in high-level mode.  The results are quite surprising: using HipCam incurs only a modest speed penalty of a maximum factor of not more than 1.5.  HipCam's memory usage is more taxing, using several gigabytes for large formalisations. The max-detail mode needs about three times as much memory as the high-level mode. 

The memory usage of HipCam's max-detail mode is higher, but 
% proportional %% Hmm... what is the proportion?
similar to the memory used by the standard proof recording approach,
allowing some inflation for the extra information used by HipCam like the shallow size.  But depending on the use of the recorded proof we can dramatically undercut these memory requirements as the high-level mode shows; this is not possible in a simple proof recording approach where larger examples would fail outright.  

%Using HipCam doesn't absolve the user from applying wrapper functions like the various variants of \texttt{hilabel}. The important difference to Tactician is that these wrapper functions are only necessary to enhance the hierarchical structure within captured proofs. Missed wrapper propagations just mean less hierarchy in the resulting hiproof, not failure to produce the hiproof. 

\section{Related Work}

In Section~1 we explained the difference between hierarchical
structure of stored proof trees and the hierachical structure of proof
script input languages (provided by languages such as
Isar~\cite{wenzel_isar_1999}).  The later may be manipulated by
text-based interfaces, for example, to fold (temporarily hide) sub-sections.

The urge to present proofs in two dimensions is widespread.  Some systems
have taken a tree-like approach from the start, using interfaces that
present proofs in a nested hierarchical form as they are developed,
enforcing structure rigidly, 
or using a GUI to build trees.  One
early example is Nuprl's tactic trees~\cite{Griffin:thesis}; another
is the Tecton system which introduced proof forests and allowed to
print out their graphical representation~\cite{kapur_overview_1994}.
The more recent ProofWeb system~\cite{kaliszyk_merging_2009} 
allows both ``flag'' style proof development as well as tree-style,
connecting each style back to source Coq code.
%% approach
%%
An interesting mix of proof scripts and a graphical
representation also appears in Hyperproof
and its methodology of heterogenous 
reasoning~\cite{barwise_heterogeneous_1993,barker-plummer_computational_2007}.
Proofscape~\cite{_proofscape_????} is a recently launched project which aims
to become a visual library of mathematics. It displays proofs with an
adjustable level of detail which corresponds to our notion of hiproof
boxes which are either collapsed or expanded in their visualisation.
%% DA: Quite a similar aim to Hiproofs, could say more perhaps

A complete survey of proof visualisation tools for proof is out of
scope, but we mention one example that inspired the
visualisation work here: the Prooftree tool by
%%% STEVEN THIS SHOULD HAVE A BETTER REF I THINK, HE HAD A PAPER
Tews~\cite{_prooftree_????} displays proof trees for Coq in Proof
General (in turn itself inspired by the similar feature provided in
PVS).  Contrary to our current visualisation software, Prooftree supports
interactive visualisation during a proof, but it does
not yet include hierarchical proof trees.

%  as an alternative to hiproofs for inducing structure on proofs which can be % exploited by the user interface. 

%% Various other systems 

%% NOT RELATED
%% I'm really not sure how Tinycals is related, I know the referee mentioned it
%% but this reads terribly
% Tinycals~\cite{coen_tinycals:_2007} represent another such formalism which allows to compose tactics in a way so that they can be stepwise executed. Tinycals work on the level of goals just as Tactician does but it is not clear how a hierarchical visualisation employing them would look like as tinycals do not fit into the hiproof framework. On the other hand, would tinycals be transplanted into a HOL Light setting, then HipCam would be immediately applicable to them, as they would have to produce valid HOL proofs. 

%% DA: I THINK THIS REFACTORING IS ANOTHER STRAND OF WORK, NOT MAIN STORY HERE
%% DA: but OK, I argued with myself and put it back in.
On a different strand, a main purpose of Tactician is to serve as a refactoring tool that can convert between ``flat'' and ``packaged up'' proofs.  There is related work on proof refactoring, including some approaches designed based on
hiproof semantics by Whiteside~\cite{whiteside_towards_2011,Whiteside:thesis}, as
well as tools that have been implemented such as the conversion between procedural and declarative proof scripts inside ProofWeb~\cite{kaliszyk_merging_2009} 
and the Levity tool~\cite{bourke12} which allows moving lemmas between different theories.  Generally such tools are still in the early days.

%  provides a general framework for refactoring of proof scripts based on hiproof semantics
% and some experimental tools.
% but it is not clear if the particular refactoring operation that Tactician provides could be studied within this framework. 
% There are not that many kinds of proof refactorings we know of that have actually been implemented. An example here is

\section{Conclusions}

Hierarchical proofs as invented in~\cite{denney_hiproofs:_2006} have so far been mostly theoretical constructs. Our work is directed towards gaining hands-on experience with ``real-world'' hiproofs. With Tactician and with HipCam we have laid the technical foundations for such an undertaking, we are able now to take existing bodies of formalisations, represent them as hiproofs, and study them as such.

Tactician can present individual proofs at the level of detail given
by the user,
% , which can be refactored and represented as hiproofs with only very modest resource usage.  
but because of its mentioned limitations, it is less suitable to
automatically obtain hiproof representations of existing large
formalisations, or to delve arbitrarily deep to understand the results
of a complicated tactic; this is what HipCam was designed for.
Tactician recovers user-level proof steps lost to HipCam, but
hierarchy doesn't appear for free in either case.  Both tools allow
the user to annotate tactics to automatically add labels (for example,
boxing up an induction or simplification tactic), or add labels
manually in particular proofs.

% 
% require the user to 
% doesn't absolve the user from applying wrapper functions like the various variants of \texttt{hilabel}, though. The difference to Tactician is that these wrapper functions are only necessary to enhance the hierarchical structure within captured proofs. Missed wrapper propagations just mean less hierarchy in the resulting hiproof, not failure to produce the hiproof. 

It is natural to ask whether the approaches can be combined.  The
disadvantage with Tactician is the need to modify scripts pervasively,
but this issue arises mainly because of its aim to record proof script
input fully to allow refactoring; without this the wrapper functions
are much simpler.  Conversely, HipCam could be provided with a
modified subgoal package like Tactician's that records user-level
proof steps and triggers only high-level capturing mode between steps.

%When viewed just as a method for capturing hiproofs, and not as a refactoring tool, it should be possible to incorporate most of the advantages of Tactician into HipCam, especially its low resource usage. The high-level capturing mode of HipCam is a first attempt at this.

More crucially, looking at hiproofs generated automatically via HipCam
from theorems like the Jordan Curve Theorem is not very illuminating,
because there isn't enough hierarchy yet.  The problem is that it is
hard to distinguish between those parts of the proof which convey its
meaning, and those parts which exist for mostly technical reasons. So
the challenge is how to ``box up'' the technical parts of a
proof, so that its meaningful parts are emphasised, and do this in a
hierarchical way.  In future work we plan to investigate
semi-intelligent ways of transforming a hiproof to introduce
structure, as well as using some manual labelling on some case study
large developments.

% Our next step will be to research if this challenge can be
% met (up to a certain extent) by applying heuristics that guide the
% automatic introduction of hierarchy / boxes.

\paragraph{Acknowledgements.}  We're grateful to the CICM referees for
providing good suggestions to improve the paper and to members of the
Mathematical Reasoning Group at Edinburgh for feedback and
discussions.  Our research was supported by funding from UK EPSRC
grant EP/J001058/1 and from the Laboratory for Foundations of Computer
Science, University of Edinburgh.

\bibliographystyle{abbrv}
\bibliography{hiproofs}
\end{document}